\documentclass[sigconf]{acmart}

\usepackage{algorithm}
\usepackage{algorithmic}
\usepackage{graphicx}
\usepackage{textcomp}
\usepackage{xcolor}
\usepackage{enumitem}
\usepackage{multirow}
\usepackage{booktabs}
\usepackage{diagbox}
\usepackage{colortbl}
\usepackage{subcaption}
\usepackage{verbatim}
\usepackage{textcomp}

\AtBeginDocument{%
  }

\setcopyright{acmlicensed}

\copyrightyear{2025}
\acmYear{2025}
\setcopyright{cc}
\setcctype{by}
\acmConference[RecSys '25]{Proceedings of the Nineteenth ACM Conference on Recommender Systems}{September 22--26, 2025}{Prague, Czech Republic}
\acmBooktitle{Proceedings of the Nineteenth ACM Conference on Recommender Systems (RecSys '25), September 22--26, 2025, Prague, Czech Republic}\acmDOI{10.1145/3705328.3748046}
\acmISBN{979-8-4007-1364-4/2025/09}

\begin{document}

\title{Exploring Scaling Laws of CTR Model for Online Performance Improvement}

\author{Weijiang Lai}
\affiliation{%
\institution{Institute of Software, Chinese
Academy of Sciences}
\institution{University of Chinese Academy of Sciences
\city{Beijing}
  \country{China}}}
\email{laiweijiang22@otcaix.iscas.ac.cn}

\author{Beihong Jin$^\dagger$}
\affiliation{%
  \institution{Institute of Software, Chinese
Academy of Sciences}
\institution{University of Chinese Academy of Sciences
\city{Beijing}
  \country{China}}}
\email{Beihong@iscas.ac.cn	}

\author{Jiongyan Zhang}
\affiliation{%
  \institution{Meituan}
  \city{Beijing}
  \country{China}}
\email{zhangjiongyan@meituan.com}

\author{Yiyuan Zheng}
\affiliation{%
  \institution{Institute of Software, Chinese
Academy of Sciences}
\institution{University of Chinese Academy of Sciences
\city{Beijing}
  \country{China}}}
\email{zhengyiyuan22@otcaix.iscas.ac.cn}

\author{Jian Dong}
\affiliation{%
  \institution{Meituan}
  \city{Beijing}
  \country{China}}
\email{dongjian03@meituan.com}

\author{Jia Cheng}
\affiliation{%
  \institution{Meituan}
  \city{Beijing}
  \country{China}}
\email{jia.cheng.sh@meituan.com}

\author{Jun Lei}
\affiliation{%
  \institution{Meituan}
  \city{Beijing}
  \country{China}}
\email{leijun@meituan.com}

\author{Xingxing Wang}
\affiliation{%
  \institution{Meituan}
  \city{Beijing}
  \country{China}}
\email{wangxingxing04@meituan.com}

\thanks{$^\dagger$ Corresponding author.}

\renewcommand{\shortauthors}{W. Lai et al.}

\begin{abstract}
Click-Through Rate (CTR) models play a vital role in improving user experience and boosting business revenue in many online personalized services. However, current CTR models generally encounter bottlenecks in performance improvement. Inspired by the scaling law phenomenon of Large Language Models (LLMs), we propose a new paradigm for improving CTR predictions: first, constructing a CTR model with accuracy scalable to the model grade and data size, and then distilling the knowledge implied in this model into its lightweight model that can serve online users. To put it into practice, we construct a CTR model named SUAN (\textbf{S}tacked \textbf{U}nified \textbf{A}ttention \textbf{N}etwork). In SUAN, we propose the unified attention block (UAB) as a behavior sequence encoder. A single UAB unifies the modeling of the sequential and non-sequential features and also measures the importance of each user behavior feature from multiple perspectives. Stacked UABs elevate the configuration to a high grade, paving the way for performance improvement. In order to benefit from the high performance of the high-grade SUAN and avoid the disadvantage of its long inference time, we modify the SUAN with sparse self-attention and parallel inference strategies to form LightSUAN, and then adopt online distillation to train the low-grade LightSUAN, taking a high-grade SUAN as a teacher. The distilled LightSUAN has superior performance but the same inference time as the LightSUAN, making it well-suited for online deployment. Experimental results show that SUAN performs exceptionally well and holds the scaling laws spanning three orders of magnitude in model grade and data size, and the distilled LightSUAN outperforms the SUAN configured with one grade higher. More importantly, the distilled LightSUAN has been integrated into an online service, increasing the CTR by 2.81\% and CPM by 1.69\% while keeping the average inference time acceptable. Our source code is available at https://github.com/laiweijiang/SUAN. 
\end{abstract}



\maketitle

\section{Introduction}
The Click-Through Rate (CTR) prediction task in online personalized services is to estimate the probability that a user will click on a desired target, such as an advertisement, a recommended item, or a search result. The CTR prediction requires not only high accuracy but also stringent latency. However, these two requirements conflict with each other, as improvements in accuracy can frequently compromise latency. Therefore, CTR models have been pursuing the goal of pushing the CTR accuracy as high as possible under the tight latency constraint. 

So far, a great deal of effort has been dedicated to developing CTR models, often improving performance by analyzing interactions between features and/or sequences of user behaviors. 
However, current CTR models seem to encounter the performance bottleneck, since in CTR prediction tasks, even a tiny increase of 0.001 in AUC (Area Under the ROC Curve) is regarded as significant \cite{twinv2, auc}, where AUC is a widely adopted metric for measuring offline accuracy. Although this viewpoint stems partly from the fact that such a minute increase can yield online gains in the industry, it highlights the difficulty in improving the performance of CTR models.

On the other hand, Large Language Models (LLMs) have been successful in various Natural Language Processing (NLP) tasks \cite{gpt3,gpt4,llama,bai2023qwen}. Extensive empirical results show that performance of LLMs improves with increasing model size and data size, a phenomenon called scaling laws \cite{kaplan2020scaling}. Recent practices also indicate that architectures that are based on specific structures, such as self-attention \cite{kaplan2020scaling, scaling_self_attention} and the mixture of experts \cite{scaling_mmone1, scaling_mmone2}, exhibit more obvious scaling laws. Meanwhile, various training and inference optimization techniques such as computation reduction and accelerated inference have been proposed to help LLMs land in different scenarios \cite{LLM_optimization_1, LLM_optimization_2, LLM_optimization_3, LLM_optimization_4, LLM_optimization_5, LLM_optimization_6}. 

Motivated by the remarkable advancements in LLMs, scaling laws have also been explored in other fields beyond NLP, such as information retrieval \cite{scaling_modal1, scaling_dense} and computer vision \cite{scaling_cv1, scaling_cv2, scaling_modal2}. Some research has explored the scaling laws in recommender systems  \cite{hstu,guo2023embedding, clue, wukong, SRT, scaling_rec}. 
However, online services do not gain substantial benefits from the models holding the scaling laws, because these models with high performance cannot be deployed in online services due to the inference time constraint.

In this paper, we present a new paradigm for improving CTR predictions, i.e., we first construct a CTR model with the scaling laws that can get better performance by increasing the model grade and data size, where the model grade, an extension of the concept of model size in the context of scaling laws of models, is defined as the model size plus length of the user behavior sequence fed into the model. This definition is grounded in the consensus that a high-performance CTR model should be able to gain more benefits from longer user behavior sequences while facing rapidly growing user behavior data.

Then, considering that high-grade CTR models have high CTR accuracy but high computational overhead that does not satisfy the inference time constraint, we apply knowledge distillation techniques to transfer the collective knowledge from a high-grade CTR model into a low-grade, more lightweight CTR model. The resulting CTR model is expected to be deployed online and serve users.

Following this line of thought, we propose a CTR model named SUAN (\textbf{S}tacked \textbf{U}nified \textbf{A}ttention \textbf{N}etwork) to offer scalable, high-accuracy CTR predictions. Specifically, we design the unified attention block (UAB), which incorporates multiple attention mechanisms to optimize representations of different features, and borrows RMSNorm and SwiGLU from LLMs to enhance training stability. 
Subsequently, we construct a LightSUAN, a lightweight version of SUAN,
and train it using the high-grade SUAN as the teacher. The distilled LightSUAN achieves a performance improvement without compromising inference time and thus can be deployed online.

Our contributions are summarized as follows: 
\begin{itemize}
\item We design the SUAN model, which comprises stacked UABs. Each UAB is equipped with multiple attention mechanisms: self-attention discerns spatiotemporal dependencies among features within the sequence; cross-attention identifies the significance of user behavior features from the user profile's perspective; and dual alignment attention selectively highlights informative features while suppressing less relevant ones.

\item We construct a LightSUAN, i.e., a SUAN with sparse self-attention and parallel inference strategies. Moreover, we adopt online distillation to train the LightSUAN and the high-grade SUAN simultaneously so that the trained LightSUAN can fully leverage the advantages of the high-grade SUAN, essentially allowing it to benefit from the scaling laws held by SUAN. 
\item We conduct extensive offline experiments on three datasets, and the experimental results show that SUAN not only has excellent AUCs compared to multiple competitors but also holds the AUCs that scale with model grade and data size spanning three orders of magnitude, and the distilled LightSUAN outperforms the original SUAN configured with one grade higher. Moreover, in the online A/B test, the distilled LightSUAN increases the CTR by 2.81\%, and CPM by 1.69\%, while keeping the average inference time acceptable. 
\end{itemize}


\section{Related Work}
\label{Related Work}

\subsection{CTR Models}
CTR models, which perform ranking tasks related to targets, have been a hot topic in both academia and industry. 

Currently, CTR models have extensively adopted deep learning technology and can be roughly divided into two categories: one focuses on feature modeling to learn the interactions between features 
\cite{wide, deepcross, xdeepfm, deepfm, dcn_v2, youtube, din, can}
, the other focuses on modeling user behavior sequences to learn user interests and/or intents \cite{dien, dsin, bst, meituan1, meituan2}.

The first category contains some classic and well-known models. For example, 
DIN \cite{din} proposes a local activation unit to model the relationship between user behavior and target item, which has a significant impact on the industry. CAN \cite{can} designs a co-action unit that parameterizes the feature embeddings in the form of a micro-MLP to fit complex feature interactions, thereby further enhancing the value of feature interactions. 

Most recent progress in CTR models falls into the second category. For example, 
DSIN \cite{dsin} introduces sessions into user behavior sequences and employs a bidirectional LSTM \cite{lstm} to learn the evolution of interests between sessions. BST \cite{bst} uses Transformer \cite{transformer} to learn user behavior sequences.
Besides, some CTR models \cite{sim, eta, twin, mimn, hpmn, ubr4ctr, sdim, long_seq1} have focused on modeling long behavior sequences. Industry practices such as SIM \cite{sim}, ETA \cite{eta}, and TWIN \cite{twin} often adopt a two-stage framework wherein long sequences are first decomposed into multiple short sequences related to the target. These short sequences are then accurately modeled in the second stage. These models can be viewed as a simplified version of a CTR model with scalability of sequence length.


\subsection{Scaling Laws in Recommender Systems}
Influenced by LLMs, researchers have begun to examine the scaling laws in recommender systems~\cite{clue,Chitlangia2023}.

For sequential recommendation tasks, LSRM \cite{LSRM} and SRT \cite{SRT} claim that they hold the scaling laws. Both of them adopt the Transformer to encode the user behavior sequences.
Specifically, LSRM proposes layer-wise adaptive dropout and switching optimizer strategies to achieve more stable training for large-scale recommendations. SRT adopts pre-training at scale and fine-tuning for downstream tasks to improve performance.

For CTR prediction tasks, simply expanding parameters (such as embedding dimension or parameters of the prediction layer) in CTR models has been found to be insufficient for scalability ~\cite{ardalani2022understanding,guo2023embedding}. To achieve the scaling laws for CTR predictions, Guo et al. \cite{guo2023embedding} propose to learn multiple embeddings for a feature, thus forming the basis of scaling up the performance. Zhang et al. \cite{wukong} propose the Wukong layer for feature interaction, which consists of a factorization machine block and a linear compress block, performing high-order and second-order feature interactions, respectively. Zhai et al. \cite{hstu} propose the HSTU architecture to model user behavior sequences for retrieval or ranking tasks.
HSTU claims that it has performance scalability w.r.t. training FLOPs whose changes arise from variations in the number of layers, sequence lengths, embedding dimensions, etc. Unfortunately, so far, the way to benefit from CTR models with scaling laws under limited inference time constraints has not been explored in depth.

Compared to existing work, our work not only proposes the key module for obtaining the scaling laws of CTR models but also provides an approach to allowing an online service to benefit from a CTR model with the scaling laws. 

\begin{figure*}[tb]
  \includegraphics[width=\textwidth]{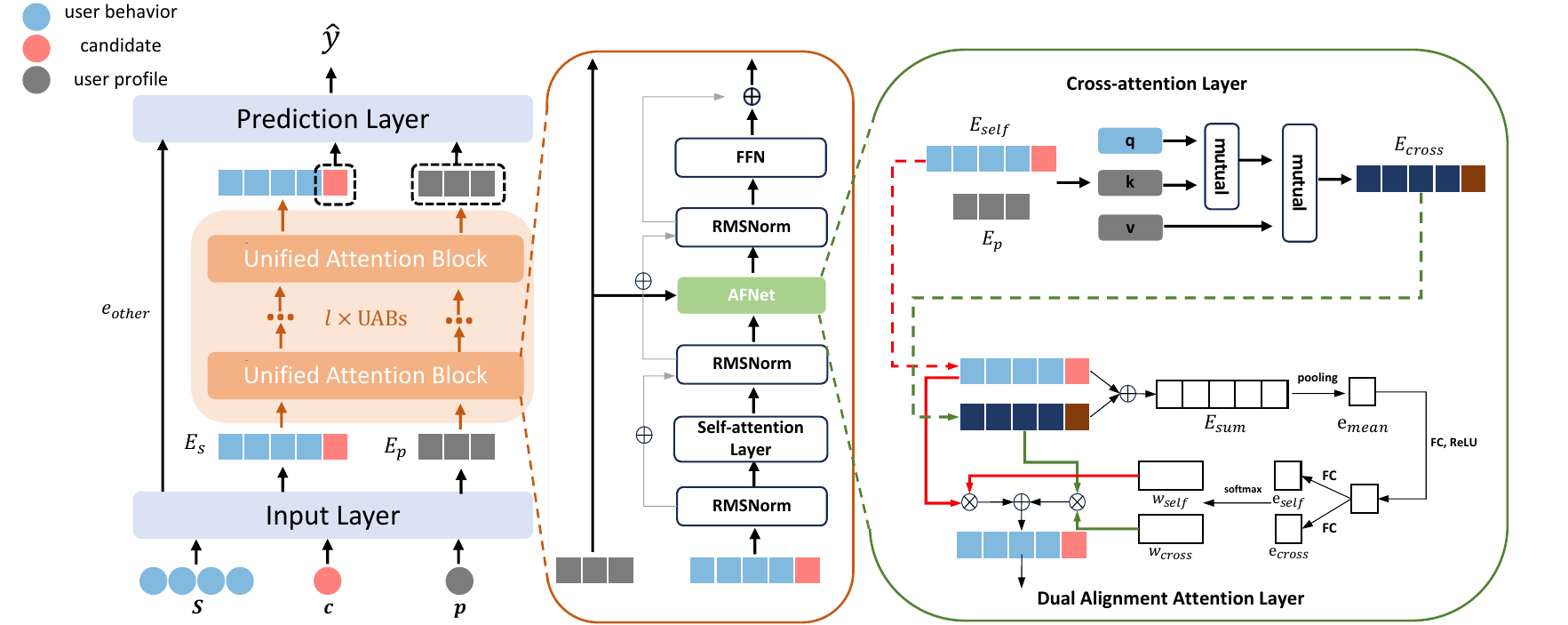}
  \caption{Architecture of SUAN.}
  \label{fig:model_architecture}
\end{figure*}

\section{Methodology}
\label{Methodology}

\subsection{Overview}
Given a user $u$, his/her behavior sequence containing $L$ behaviors is denoted by $S={\{b_i\}}_{i=1}^L$, where $b_i$ denotes the $i$-th behavior, including features such as item ID and timing of behavior. His/Her user profile is denoted by $p$, with features such as user age and gender. For a candidate $c$ w.r.t. the user $u$, it includes regular features such as item ID, exposure time, and detailed features such as its CTR of the last 3 days. 

Our goal is to build a model that predicts the probability of user $u$ clicking $c$, that is, $\mathcal{P}(c | S,p)=F(S,p,c; \theta)$, where $F$ denotes our model, and $\theta$ denotes the model parameters. 

For this goal, we propose a model named SUAN, as illustrated in Figure~\ref{fig:model_architecture}, which consists of an input layer, $l$ UABs as encoders and a prediction layer, where $l$ is the number of UABs.

Since item IDs are numerous, sparse, and often become invalid or outdated compared to NLP tokens, we concatenate regular features of behaviors, such as item ID and category ID, to feed into the input layer. This approach alleviates sparsity and provides collaborative information, enabling better modeling of user interests.
Additionally, we append the candidate’s regular features to the user behavior sequence to form a target-aware sequence, still using $L$ to denote length for simplification. It helps the encoder understand the relationships between user behaviors and the candidate, enabling the model to focus on behaviors more relevant to the candidate, which is highly beneficial for CTR tasks.

In the input layer, we use a uniform embedding table to initialize target-aware sequence embedding matrix $E_s \in \mathbb{R}^{L \times n_1d}$, user profile embedding matrix $E_p \in \mathbb{R}^{n_2 \times d}$ and other detailed features embedding vector $e_{other} \in \mathbb{R}^{n_3d}$, where $n_1,n_2,n_3$ are the numbers corresponding to features, and $d$ is the dimension.



Within the $l$ UABs, each block models the target-aware sequence via various attention mechanisms, capturing complex relationships and patterns within the sequence, as detailed in Section~\ref{UAB}.

In the prediction layer, the candidate representation from the final UAB's output $E_{block}[-1,:]$, along with the vector $e_p$ obtained by flattening $E_p$ and other features $e_{other}$ are fed into the multi-layer perception (MLP), and the probability of clicking the candidate $\hat y$ is predicted as follows:
\begin{equation}
\label{equation2}
    \hat{y} = \sigma(z), \quad z = \mathrm{MLP}(E_{\mathrm{block}}[-1,:], e_p, e_{\mathrm{other}}),
\end{equation}
where $\sigma$ is the sigmoid function and $z$ denotes the logits.


We employ the standard binary cross-entropy loss, denoted as $L_{ce}(\hat{y}, y)$, to optimize our model, where $y$ denotes the ground truth of $\hat{y}$.





\subsection{Unified Attention Block}
\label{UAB}
We propose the UAB as a causal encoder, adopting self-attention as its backbone.

At the start of UAB, we apply a pre-norm strategy to normalize the target-aware sequence embedding $E_s$, as follows:
\begin{equation}
    E_{\mathrm{norm}} = \mathrm{RMSNorm}(E_s),
\end{equation}
where RMSNorm \cite{rmsnorm} is a variant of layer normalization that utilizes only the root mean square for layer normalization.  
This strategy, commonly used in LLMs, is introduced into CTR prediction for the first time, serving as a regularizer to reduce computation and maintain training stability.

A complete UAB, in addition to the RMSNorm, includes the self-attention layer, adaptive fusion network (AFNet), and feedforward network (FFN) to discover the importance of features from various perspectives, thereby enhancing the model's expressiveness.

\subsubsection{Self-attention Layer}
To model target-aware sequences, we use the self-attention mechanism to identify key behaviors, thereby capturing user interests. Specifically, we add the attention bias \cite{attentionbias, hstu} to the scaled dot-product attention scores
to learn the relative positional and temporal relationships between elements. In this layer, we feed with the normalized target-aware sequence embedding $E_{norm}$ to obtain the self-augmented sequence embedding $E_{self}$ as follows:
\begin{equation}
    Q_1, K_1, V_1 = E_{\mathrm{norm}}W_1^Q, E_{\mathrm{norm}}W_1^K, E_{\mathrm{norm}}W_1^V,
\end{equation}
\begin{equation}
    \mathrm{bias}_{ij} = f_1(\Delta t_{ij}) + f_2(\Delta p_{ij}),\ i, j \in \{1, 2, \ldots, L\},
\end{equation}
\begin{equation}
    \mathrm{Attention}(Q_1, K_1, V_1) = \mathrm{softmax}\left(\frac{Q_1 K_1^T}{\sqrt{n_1d}} + \mathrm{bias}\right)V_1,
\end{equation}
\begin{equation}
    E_{\mathrm{self}} = \mathrm{Concat}(\mathrm{Attention}_1, \ldots, \mathrm{Attention}_h)W_1^O,
\end{equation}
where $W_1^Q,W_1^K,W_1^V$, $W_1^O \in \mathbb{R}^{n_1 d \times n_1 d}$ are projection matrices. $\Delta t_{ij}$ and $\Delta p_{ij}$ donate the relative time and relative position difference between the $i-$th and the $j-$th elements, respectively, and $f_1 (\cdot),\ f_2 (\cdot)$ are mapping functions that project the difference into learnable variables. $h$ is the number of attention heads.
Since there are no obviously temporal or associative relationships among user profile features, such as user age and gender, we opt not to encode them using the self-attention layer for efficiency.

\subsubsection{Adaptive Fusion Network}
To facilitate effective interaction between sequential and non-sequential features while maintaining feature space consistency, we propose the AFNet, which uniformly models user sequential and non-sequential features. Specifically, AFNet includes a cross-attention layer and a dual alignment attention layer. The cross-attention layer learns the relationship between sequential and non-sequential features to generate the cross-augmented sequence embedding. 
The dual alignment attention adaptively aligns self-augmented sequence embedding and cross-augmented sequence embedding from the perspective of the global distribution of features.

\noindent\textbf{Cross-attention layer.} 
In this layer, we design unidirectional cross-attention to analyze user behaviors, with the intention of diminishing noise and highlighting key behaviors from the perspective of the user profile, thereby enhancing the expressiveness of sequence embeddings. 

This approach is based on the fact that user profile features typically offer a more accurate representation of the user and effectively reflect their interests than behavior features. Thus, incorporating behavior features into the user profile representation tends to introduce noise, affecting its accuracy. Conversely, the user profile can provide precise information for refining user interests. Specifically, we take the self-augmented sequence embedding $E_{self}$ as query and user profile embedding $E_p$ as key and value to obtain the cross-augmented sequence embedding $E_{cross}$, as shown below:
\begin{equation}
    Q_2, K_2, V_2 = E_{\mathrm{self}}W_2^Q, E_pW_2^K, E_pW_2^V,
\end{equation}
\begin{equation}
    \mathrm{Attention}(Q_2, K_2, V_2) = \mathrm{softmax}\left(\frac{Q_2 K_2^T}{\sqrt{n_1d}}\right)V_2,
\end{equation}
\begin{equation}
    E_{\mathrm{cross}} = \mathrm{concat}(\mathrm{Attention}_1, \ldots, \mathrm{Attention}_h)W_2^O,
\end{equation}
where $W_2^Q, W_2^O \in \mathbb{R}^{n_1d\times n_1d}, W_2^K, W_2^V \in \mathbb{R}^{d\times n_1d}$ are projection matrices. 

\noindent\textbf{Dual Alignment Attention Layer.} 
The self-augmented sequence embedding $E_{self}$ and the cross-augmented sequence embedding $E_{cross}$ aggregate internal behavior relationships and external profile semantics, respectively. In this layer, we combine them to capture comprehensive user interests. For each behavior, the importance of other behaviors and profile contexts differs, making it essential to learn the significance of individual self-augmented and cross-augmented features. Inspired by the channel-wise attention\cite{senet1, senet2}, we propose dual alignment attention to align user behavior and profile information for each behavior adaptively. We first generate dimension-wise statistics $e_{mean}$ using global mean pooling as follows:
\begin{equation}
    E_{sum}=E_{self}+E_{cross},
\end{equation}
\begin{equation}
    e_{mean}=\frac{1}{L}\sum_{i=1}^L E_{sum}(i,:).
\end{equation}

Furthermore, we utilize two fully connected (FC) layers to implement the gating mechanism: the first layer suppresses less useful features, and the second layer selectively emphasizes informative features, as demonstrated in the following formulas:
\begin{equation}
    e_{self},e_{cross}=(\rho(e_{mean}W_1))W_2,(\rho(e_{mean}W_1))W_3,
\end{equation}
\begin{equation}
    w = \mathrm{softmax}\left(\left[\begin{array}{c} e_{self} \\ e_{cross} \end{array}\right]\right) \in \mathbb{R}^{2 \times L \times n_1d},
\end{equation}
\begin{equation}
    w_{self},w_{cross}=w[0,:,:],w[1,:,:],
\end{equation}
where $\rho$ is ReLU activation function, and $W_1 \in \mathbb{R}^{n_1d \times \frac{n_1d}{4}}, W_2, W_3 \in \mathbb{R}^{\frac{n_1d}{4} \times n_1d}$. Next, we obtain the comprehensive user sequence embedding $E_{AFN}$ by the following formula:
\begin{equation}
    E_{AFN} = w_{self}\odot E_{self}+w_{cross}\odot E_{cross}.
\end{equation}

In short, in this layer, we dynamically integrate self-augmented sequence embedding and cross-augmented sequence embedding by perceiving the relationships between user behavior and user profile, and the global feature distribution in the sequence.

\subsubsection{Feedforward Network}
We adopt SwiGLU-based feedforward network (FFN), which is formulated as follows:
\begin{equation}
    E_{FFN}=(\phi (E_{AFN}W_1^{FFN}) \odot E_{AFN}W_2^{FFN})W_3^{FFN},
\end{equation}
where $\phi$ is the Swish activation function \cite{swish}, $W_1^{FFN}, W_2^{FFN} \in \mathbb{R}^{n_1d \times 3n_1d}, W_3^{FFN} \in \mathbb{R}^{3n_1d \times n_1d}$. 
The FFN combines the property of being differentiable everywhere in the Swish function with the advantage of the gating mechanism in GLU~\cite{glu}. Similar FFNs are used in LLMs.





\subsection{Deployment Optimization}
\subsubsection{Speedup Strategies}
To alleviate the high time complexity of the SUAN model, we adopt sparse self-attention and parallel inference strategies, and the modified model is referred to as LightSUAN.

\noindent\textbf{Sparse self-attention.} 
As we know, the time complexity of the self-attention mechanism is quadratic in the sequence length, which can significantly affect model training and inference times when the sequence length is long. Sparse self-attention seeks to lessen the computational burden by reducing the number of attention connections. 

\begin{figure}[t]
  \centering
  \begin{minipage}[t]{0.48\textwidth}
    \centering
    \includegraphics[width=\linewidth]{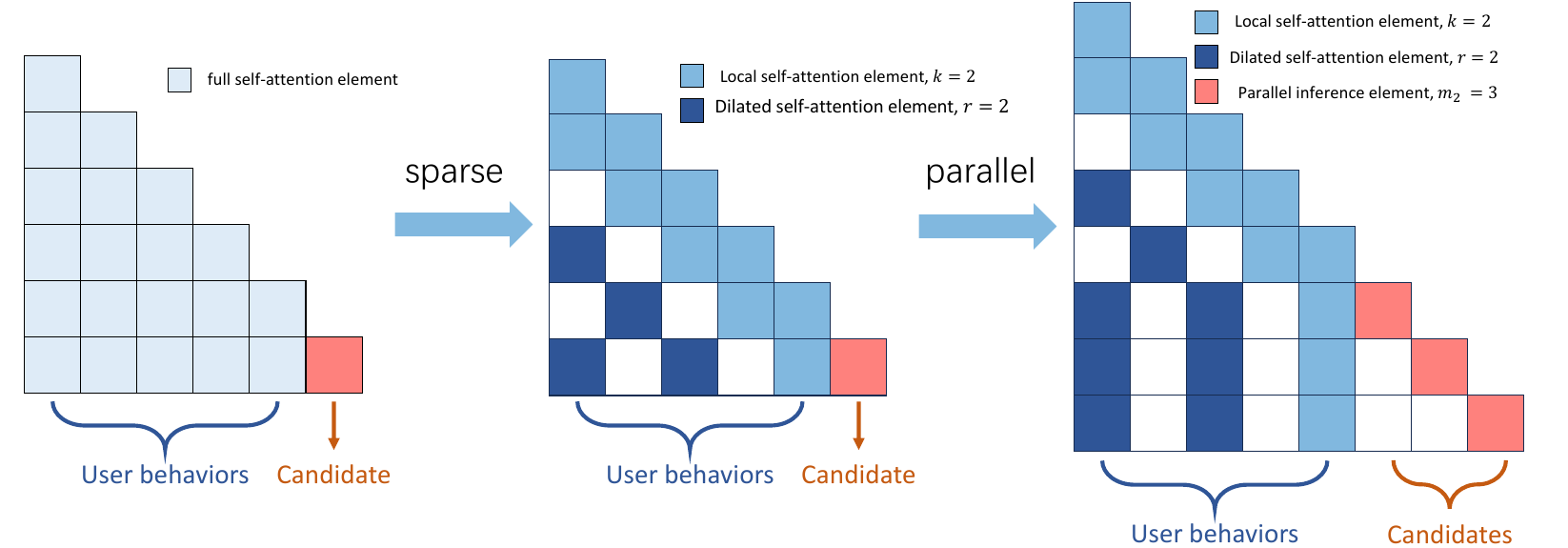}
    \caption{Illustration of self-attention matrix in LightSUAN.}
    \label{fig:sparse}
   \end{minipage}
 \hfill 
 \begin{minipage}{0.48\textwidth}
  \centering
  \includegraphics[width=0.9\linewidth]{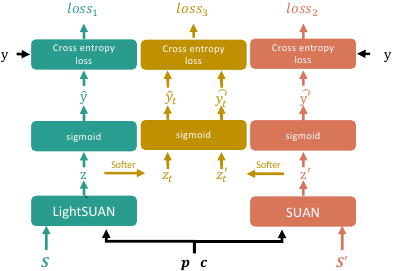}
  \caption{Illustration of online distillation.}
  \label{fig:distill}
 \end{minipage}
\end{figure}

Our sparse self-attention is a combination of local self-attention and dilated self-attention, enabling the effective capture of both recent and distant relationships~\cite{sparse2}. In the local self-attention, each element attends only to its neighboring elements within a local window, where $k$ denotes the window size. In the dilated self-attention, each element attends only to every $r$-th element, where $r$ is the dilated rate. The middle part of Figure~\ref{fig:sparse} shows the method of calculating a sparse self-attention matrix, where $r=2$ and $k=2$.

\noindent\textbf{Parallel inference.} 
For a user, online CTR prediction involves ranking multiple candidates. If there are $m_1$ candidates, then $m_1$ separate inferences are performed conventionally. Since CTR models have strict requirements for inference time, it is essential to propose a parallel inference strategy to accelerate online inference.

In SUAN, our causal encoder ensures that behavior representations are independent of candidate representations, which enables parallel inference through the reuse of behavior representations. 
Our parallel inference strategy requires modifications to online inference and offline training. For online inference, since the value of $m_1$ is variable in a specific online service, we input $m_2$ $(m_2 < m_1)$ candidates for parallel inference, obtaining $m_2$ results in a separate inference. This reduces the number of inferences to $\lceil{m_1}/{m_2}\rceil$ per user. To match the online parallel inference, we modify the implementation specifics of offline training to simulate the online scenario. First, we modify the training samples to ensure consistency with the input of the online inference. To be specific, for an original training sample that includes only a single candidate, we append $m_2-1$ placeholders to this candidate. Then, we modify the method of calculating the attention matrix. The right part of Figure~\ref{fig:sparse} shows an example illustrating how to perform parallel inference in conjunction with the sparse self-attention mechanism, where $k=2$, $r=2$, and $m_2=3$.


\subsubsection{Online distillation}

To efficiently transfer knowledge from a high-grade SUAN to a deployable low-grade LightSUAN, we design an online distillation approach that trains these two models simultaneously after examining various distillation techniques~\cite{ distilling1, distilling2, online_distilling1, online_distilling2, online_distilling3}.

As illustrated in Figure~\ref{fig:distill}, a low-grade LightSUAN acts as a student, fed with $S, c, p$, and a high-grade SUAN acts as a teacher, fed with $S', c, p$. $\hat y$ and $\hat {y'}$ denote the predication probabilities and $z$ and $z'$ denote the logits output by two models. Besides, $z_t=z/t$, $z_t' = z'/t$ where $t$ denotes a temperature, resulting in softer probability distributions over classes. Further, $\hat{y_t} =\sigma(z_t), \hat{y_t'}=\sigma(z_t')$. 
We optimize two models using three binary cross-entropy loss functions: $loss_1 = L_{ce}(\hat y, y)$, $loss_2= L_{ce}(\hat{y'},y)$, and $loss_3=L_{ce}(\hat{y_t},\hat{y_t'})$, where $\hat{y_t'}$ acts as a soft label.



The final loss for the online distillation is as follows:
\begin{equation}
loss = loss_1+loss_2+\lambda loss_3,
\end{equation}
where $\lambda$ is the weight coefficient for the distillation loss. Since magnitudes of the gradients produced by $loss_3$  scale as $1/t$ \cite{distilling1}, we set $\lambda$ to $t$ to counterbalance this scaling, ensuring that $loss_3$ contributes appropriately to the overall gradient during the optimization process.


\subsection{Complexity Analysis}
\subsubsection{Time complexity}
SUAN's time complexity primarily comes from the self-attention and cross-attention layers within the UAB. The primary complexity for $l$ UABs with full attention is $O (BlL^2n_1d + BlLn_2n_1d)$, where $B$ is the batch size.

In LightSUAN, we adopt the sparse self-attention to reduce the time complexity to $O(BlL(k+\frac{L}{r})n_1d+BlLn_2n_1d)$. Furthermore, we use the parallel inference strategy to decrease the inference time per user from $O(m_1(BlL(k+\frac{L}{r})n_1d+BlLn_2n_1d))$ to $O(\frac{m_1}{m_2}(Bl(L+m_2)(k+\frac{L+m_2}{r})n_1d+Bl(L+m_2)n_2n_1d))$.

\subsubsection{Space complexity}
In addition to embeddings, the extra learnable parameters in our model primarily come from the UABs, which include the self-attention layer, AFNet, and FFN, their learnable parameters are $O(4ln_1^2d^2)$, $O(l(2n_1^2d^2+2n_1n_2d^2+\frac{3n_1^2d^2}{4}))$ and $O(9ln_1^2d^2)$, respectively.

\begin{table}[t]
\centering
\caption{Statistics of the datasets.}
\setlength\tabcolsep{4.2pt}
\begin{tabular}{ccccccc}
\hline
Dataset        & \#Users & \#Items & \#Samples \\
\hline
Eleme  & 14,427,689   & 7,446,116   & 128,000,000         \\
Taobao  & 1,141,729   &  461,527  & 700,000,000         \\
Industry         & 88,976,000   & 14,054,691  & 1,230,715,133   \\
\hline
\end{tabular}
\label{tab:datsets}
\end{table}

\section{Experiments}
\label{Experiments}

\begin{table*}[tbp]
  \caption{Performance comparison. We repeat each experiment three times and report the average results. We also report the standard deviation of SUAN's results. In each row, the best and second-best results are highlighted in bold, and the third-best results are underlined. DIN is considered the base model for calculating the RelaImpr.}
   \label{tab:performance_comparison}
  \centering
\begin{tabular}{lc|cc|cccc|cc|cc}
    \toprule
    \multirow{2}{*}{Dataset} & \multirow{2}{*}{Metric} & \multicolumn{2}{c|}{Group \uppercase\expandafter{\romannumeral1}} & \multicolumn{4}{c|}{Group \uppercase\expandafter{\romannumeral2}} & \multicolumn{2}{c|}{Group \uppercase\expandafter{\romannumeral3}} \\
    & & DIN & CAN & SoftSIM & HardSIM & ETA & TWIN & BST & HSTU & SUAN & SUAN(L)\\
    \midrule
    \multirow{2}{*}{Industry} & AUC & 0.7002 & 0.7004 & 0.7025 & 0.7020 & 0.7024 & 0.7028 & 0.7028 & \underline{0.7036} & \textbf{0.7098}$\pm$0.00004 & \textbf{0.7135}$\pm$0.00048\\
    & RelaImpr &0.00\% & 0.10\% & 1.15\% & 0.90\% & 1.10\% & 1.30\% & 1.30\% & \underline{1.70\%}  & \textbf{4.80\%}$\pm$0.02\% & \textbf{6.64\%}$\pm$0.24\%\\
    \midrule
    \multirow{2}{*}{Eleme} & AUC & 0.6363 & 0.6378 & 0.6399 & 0.6389 & 0.6398 & 0.6410 & 0.6600 & \underline{0.6631} & \textbf{0.6669}$\pm$0.00028 & \textbf{0.6690}$\pm$0.00090 \\
    & RelaImpr & 0.00\% & 1.10 \% & 2.64\% & 1.90\% & 2.56\% & 3.44\% & 17.38\% & \underline{19.66\%} & \textbf{22.45\%}$\pm$0.21\% & \textbf{23.99\%}$\pm$0.66\% \\
    \midrule
    \multirow{2}{*}{Taobao} & AUC & 0.6198 & 0.6184 & 0.6212 & 0.6239 & 0.6220 & 0.6215 & 0.6370 & \underline{0.6397} & \textbf{0.6472}$\pm$0.00011 & \textbf{0.6495}$\pm$0.00088\\
    & RelaImpr & 0.00\% & -1.17\% & 1.17\% & 3.42\% & 1.84\% & 1.42\% & 14.36\% & \underline{16.61\%} & \textbf{22.87\%}$\pm$0.09\% & \textbf{24.97\%}$\pm$0.73\%\\
    \bottomrule
\end{tabular}
\end{table*}

\subsection{Experimental Settings}
\subsubsection{Datasets}
We adopt two public datasets and one industrial dataset to conduct experiments. \textbf{Eleme}\footnote{https://tianchi.aliyun.com/dataset/131047} is constructed from logs of the Ele.me service and contains 30-day behaviors of users. It has abundant behavior features, including item features and behavior features. 
\textbf{Taobao}\footnote{https://tianchi.aliyun.com/dataset/56
} is collected from the display advertising system in Alibaba and contains 22-day behaviors of nearly a million randomly chosen users on Taobao. Each user behavior has a behavior type, the time when behavior has occurred, and so on.
\textbf{Industry} is an industrial dataset that contains up to 1000 behavioral records from the past year for each of 88 million users who are sampled from active users during the 7-day period from July 6, 2024, to July 12, 2024. Each user behavior includes item features and behavior features. 
The statistics of three datasets are shown in Table~\ref{tab:datsets}.


\subsubsection{Competitors}
We choose eight methods of different research lines as competitors. Two of them, i.e., \textbf{DIN} \cite{din} and \textbf{CAN} \cite{can}, focus on modeling behavior feature interactions (assigned to Group \uppercase\expandafter{\romannumeral1}), and the other six, i.e., \textbf{SoftSIM}, \textbf{HardSIM}, \textbf{ETA} \cite{eta}, \textbf{TWIN} \cite{twin}, \textbf{BST} \cite{bst} and \textbf{HSTU} \cite{hstu}, adopt behavior sequence modeling, where \textbf{SoftSIM} and \textbf{HardSIM} are two variants of \textbf{SIM} \cite{sim}, using the embedding similarity retrieval and category retrieval, respectively. Of these six methods, the first four are designed for long sequences (assigned to Group \uppercase\expandafter{\romannumeral2}), and the last two are not (Group \uppercase\expandafter{\romannumeral3}). 


\subsubsection{Evaluation Metrics}
In the offline experiments, we adopt AUC as an evaluation metric.
Besides, we follow \cite{relaimpr1, din} to introduce the relative improvement (Relalmpr) metric to measure relative improvement between models.
In the online A/B test, we adopt CTR, Cost Per Mille (CPM) and inference time as evaluation metrics.

\subsubsection{Implementation Details}
\label{subsec:Implementation_Details}
We implement all models by TensorFlow. For the sake of fairness, all models are configured to have parameters of the same order of magnitude. Specifically, the dimension of all models is set to 8, and the same prediction layer was set for all models, i.e., the MLP structure in the prediction layer is [1024, 512, 256, 1], using the Dice as the activation function. Meanwhile, for our model and the models in Group \uppercase\expandafter{\romannumeral3}, we set their encoder layers to 2 and the attention heads to 2. During distillation, the hyperparameters $t$ and $\lambda$ are set to 2.

In addition, for the industrial dataset and Taobao dataset, we set the sequence length to 1000 for the models in Group \uppercase\expandafter{\romannumeral2} and 100 for the models in Groups \uppercase\expandafter{\romannumeral1} and \uppercase\expandafter{\romannumeral3}. Due to limitations of the Eleme dataset, we set the sequence length to 50 for the models in Group \uppercase\expandafter{\romannumeral2} and 10 for the models in Groups \uppercase\expandafter{\romannumeral1} and \uppercase\expandafter{\romannumeral3}. Moreover, in addition to SUAN, which is set to the same sequence length as the models in Group \uppercase\expandafter{\romannumeral1}, we also employ SUAN(L), i.e., the SUAN with the same sequence length as the models in Group \uppercase\expandafter{\romannumeral2}, to evaluate the performance of our model on long sequences.


We train all models on NVIDIA A100-80G. We train each of them for one epoch, using Adam as the optimizer.

\subsection{Overall Performance}
The performance evaluation results are listed in Table~\ref{tab:performance_comparison}. 
We also conduct a $t$-test on the AUCs of our model and each comparison model, while setting a significance level to 0.05. All $p$-values are less than 0.05, indicating a statistically significant difference between our model and the other models in terms of AUC. 

From the results, we find that the CTR models for long sequences outperform DIN and CAN, indicating that incorporating longer user behaviors helps learn more comprehensive user interests, which is beneficial for CTR predictions. Compared to the CTR models for long sequences, the CTR models in Group  \uppercase\expandafter{\romannumeral3} 
and SUAN take shorter user behavior sequences as input, but perform better. There likely exist two reasons. First, the models in Group \uppercase\expandafter{\romannumeral2}  decompose a long sequence into short subsequences, potentially losing valuable information that is implied in the long sequence. Second, the models in Group \uppercase\expandafter{\romannumeral3} and SUAN utilize the self-attention mechanism to analyze user behaviors and learn the relationships between those behaviors, thereby gaining a better understanding of user interests.


Last and most importantly, our model significantly outperforms all competitors on three datasets. The results from column SUAN(L) also demonstrate that our model can further enhance its performance by taking longer sequences as input. 

\begin{table}[t]
\centering
\caption{Results of the ablation study.}
\begin{tabular}{c|ccc}
\hline
Model        & Industry & Eleme & Taobao \\
\hline
SUAN         & 0.7098  & 0.6669  & 0.6472         \\
Variant A  & 0.7093 &0.6644  & 0.6468           \\
Variant B & 0.7075 &0.6644 & 0.6431 \\
Variant C & 0.7080 &0.6652 & 0.6441 \\
Variant D & 0.7042 & 0.6638 & 0.6434 \\
Variant E & 0.7040 &0.6631 & 0.6433 \\
\hline 
\end{tabular}
\label{tab:ablation}
\end{table}

\subsection{Ablation Study }
\label{Ablation}
We construct five variants to evaluate the effectiveness of the modules in SUAN. Variants A, B are the models that remove SwiGLU and AFNet from UAB, respectively. Variant C is to replace dual alignment attention in AFNet with a simple strategy that directly combines the self-augmented sequence embedding and cross-augmented sequence embedding using addition. Variant D is the model that replaces the target-aware sequence in SUAN with a pure user behavior sequence and is directly fed with the candidate's feature representations in the prediction layer. In other words, variant D abandons the target-aware sequence in SUAN. 
Variant E simplifies UAB by removing the attention bias of the self-attention layer.

The results are shown in Table~\ref{tab:ablation}. Compared to SUAN, all these variants exhibit varying degrees of performance degradation, illustrating the effectiveness of the designed modules.
Especially, variant D shows substantial performance deterioration, highlighting that the target-aware sequence, which exploits the distinctiveness of the CTR prediction task, plays a significant role in improving performance. 
Variant E has the greatest decline in performance, indicating that the relative position and temporal information between behaviors are crucial for learning relationships among behaviors.

\begin{figure*}[t]
    \centering
    \begin{subfigure}{.33\textwidth}
        \centering
        \includegraphics[width=\linewidth]{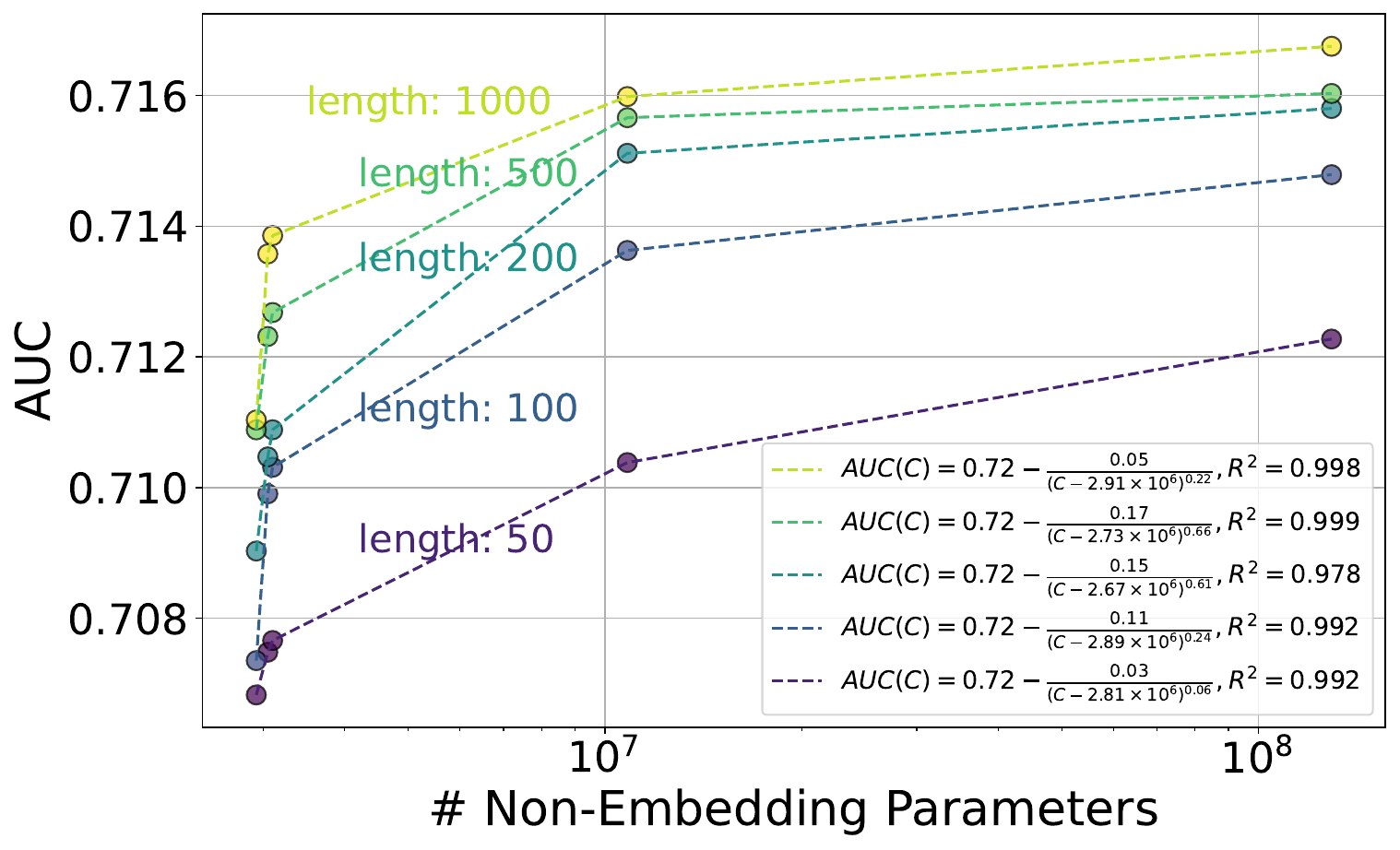}
        \caption{}
        \label{fig:auc_modelsize}
    \end{subfigure}%
    \begin{subfigure}{.33\textwidth}
        \centering
        \includegraphics[width=\linewidth]{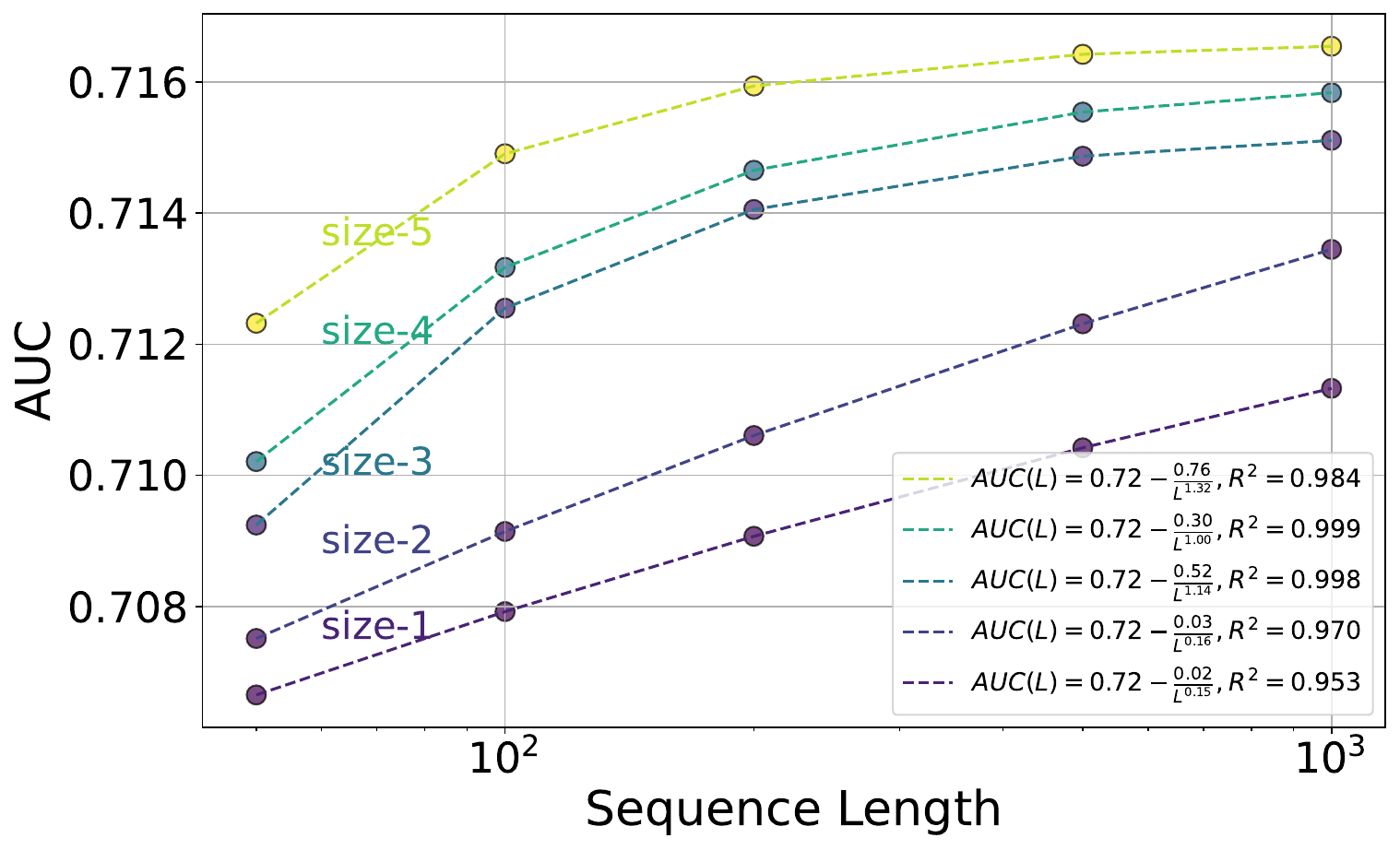}
        \caption{}
        \label{fig:auc_length}
    \end{subfigure}
    \begin{subfigure}{.33\textwidth}
        \centering
        \includegraphics[width=\linewidth]{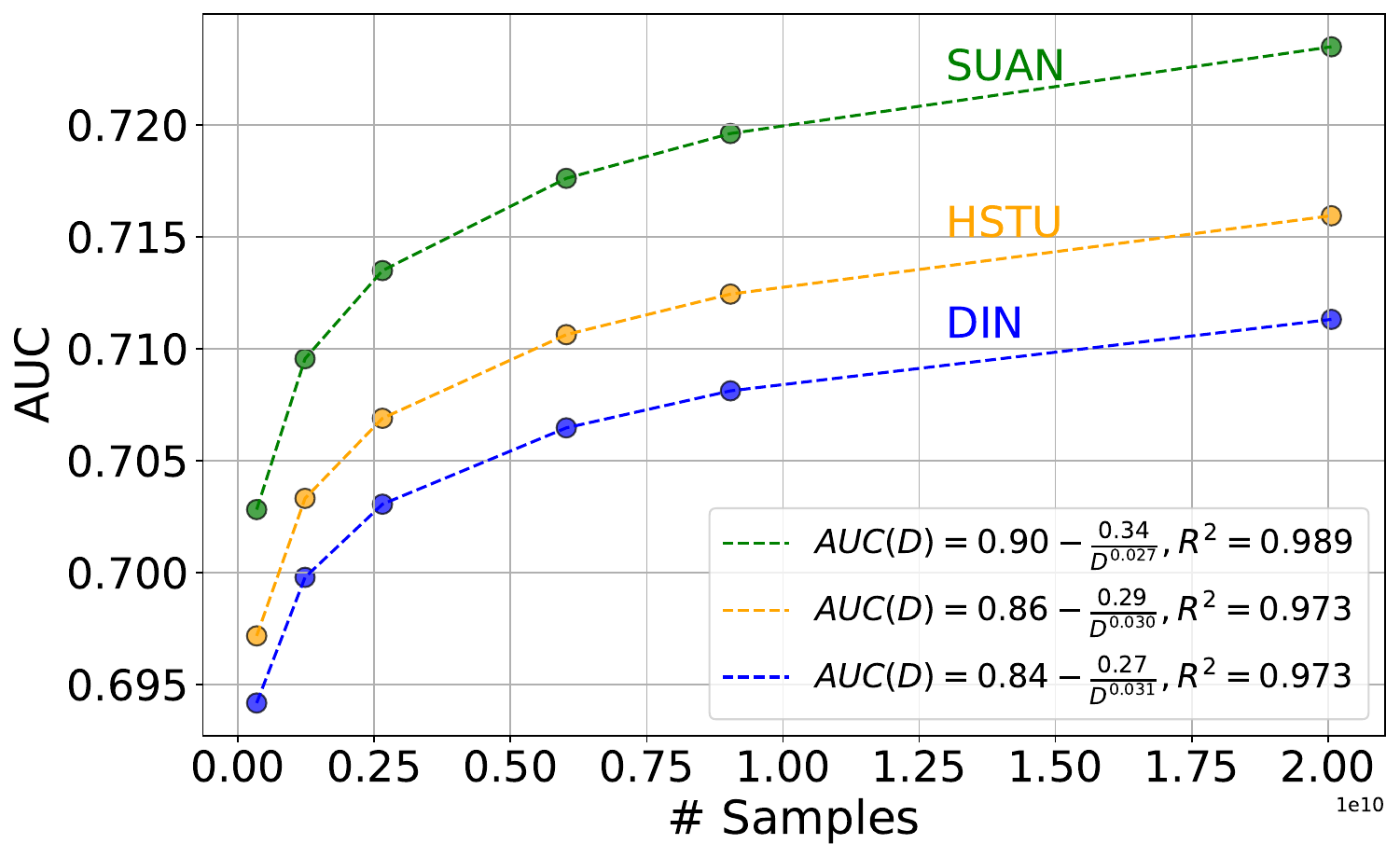}
        \caption{}
        \label{fig:auc_datasize}
    \end{subfigure}
    \caption{Scaling laws on the industrial dataset: (a) Model sizes, (b) Sequence lengths, and (c) Data size.}
    \label{fig:auc_modelsize_length_datasize}
\end{figure*}

\subsection{Scaling Law Analysis}
Similar to existing studies on scaling laws \cite{kaplan2020scaling,scaling_mmone1, scaling_mmone2, mmoe}, the way to determine whether the scaling law phenomenon exists in our model is by examining the effect of model grade and data size on model performance. In SUAN, the model grade is determined by two factors: the model size and behavior sequence length, where model size typically refers to the number of non-embedding parameters, which are primarily influenced by the dimension $d$, and the parameters $l$ and $h$ of the UBA in SUAN. The data size refers to the number of samples that are input into the model. Unlike existing studies focusing on the loss values \cite{kaplan2020scaling, LSRM}, our exploration is more concerned with AUC, as this metric is more valuable than the loss value in CTR prediction scenarios. Limited by the data size of public datasets, we conduct experiments on the industrial dataset.

\subsubsection{Impact of Scaling the Model Grade}
We take 7-day samples as input, and run SUAN under different combinations of model size and sequence length, where the model size is set according to Table~\ref{tab:scale} and the sequence length is selected from \{50, 100, 200, 500, 1000\}. For each combination, we calculate the AUC.

Taking inspiration from \cite{scaling_law_funcion, SRT}, we propose fitting the AUCs into a power-law function $    AUC(C)=E_1- {A_1}/{(C-B_1)^\alpha}$, where $C$ is the number of non-embedding parameters, $E_1$, $A_1$, $B_1$, and $\alpha$ are the coefficients to be fitted. $E_1$ can be explained as the potential upper bound of AUC values. $B_1$ can be understood as the number of parameters not tied to the scaling laws in the CTR prediction task, such as the parameters in the prediction layer. We apply the least squares method to fit AUCs into a nonlinear curve. The results are displayed in Figure~\ref{fig:auc_modelsize}. The coefficients of determination ($R^2$) are close to 1, which indicates that the fits are of high quality. From the results, we confirm that, while fixing data size, AUCs of SUAN adhere to a power-law function with respect to model size across different sequence lengths.

Much like AUC(C), we employ $AUC(L)=E_2-{A_2}/{L^\beta}$ to investigate the relationship between the AUC value and sequence length, where $L$ denotes the sequence length, and $E_2, A_2$, and $\beta$ are the coefficients.  
The fitted curves and formulas are shown in Figure~\ref{fig:auc_length}, confirming that when given fixed-size data as input, a scaling law holds when the sequence length varies. 

From the results of AUC(C) and AUC(L) in Figure~\ref{fig:auc_modelsize_length_datasize}, we find that the values of $E_1$ and $E_2$ are both approximately 0.72, indicating that an upper bound does exist for the AUC improvement under a fixed data size.

\begin{table}
\centering
\caption{Configurations of model sizes.}
\setlength\tabcolsep{4.2pt}
\begin{tabular}{c|ccc|c}
\hline
\multirow{2}{*}{Model Size} & \multirow{2}{*}{$d$} & \multirow{2}{*}{$l$} & \multirow{2}{*}{$h$} & \multirow{2}{*}{\shortstack{\#Non-embedding \\ Parameters}} \\
& & & & \\  
\hline
size-1  & 8  & 1  & 1  & 2,929,118       \\
size-2  & 8  & 2  & 2  & 3,050,358       \\
size-3  & 16  & 4  & 4  & 4,102,838       \\
size-4  & 32 & 8  & 8  & 10,824,646      \\
size-5  & 144 & 12 & 12 & 129,363,030     \\
\hline 
\end{tabular}
\label{tab:scale}
\end{table}

\subsubsection{Impact of Scaling the Data Size} 
We choose three models, i.e., SUAN, HSTU, and DIN, whose configurations are identical to those in Section~\ref{subsec:Implementation_Details}. For each model, we fix its configuration and run the model using different numbers of samples as input. Here, samples are collected from \{3, 7, 14, 30, 45, 90\} days, and the number of samples spans three orders of magnitude from 3.43 $\times 10^8$ to $2 \times 10^{10}$. We calculate the AUC for each run of the model, and then fit AUCs using $AUC(D)=E_3- {A_3}/{D^\gamma}$, where $D$ is the number of samples, and $E_3$, $A_3$, and $\gamma$ are the coefficients to be estimated. The curves and formulas are shown in Figure~\ref{fig:auc_datasize}. We find that the power-law function of SUAN exhibits a higher $E_3$ and a smaller $\gamma$ value compared to HSTU and DIN. This implies that our model can utilize training samples more effectively and exhibits excellent scalability across various data scales.

Furthermore, we find that $E_3$ values are higher than $E_2$ and $E_1$ values, and $\gamma$ values are lower than both $\alpha$ and $\beta$ values. This means that increasing the data size, as opposed to the model grade, achieves a higher upper bound and a faster growth rate of AUC for our model. 

\subsubsection{Regularizers for Supporting Scaling Laws}
\begin{table}
    \centering
        \caption{Performance of models with different model sizes.}
    \begin{tabular}{c|cc}
        \hline
        Model & size-2 & size-4 \\
        \hline
        SUAN & 0.7098±0.00004 & 0.7133±0.00128 \\
        Variant F & 0.7092±0.00068 & 0.7097±0.00331 \\
        Variant G & 0.7097±0.00019 & 0.7064±0.00463 \\
        \hline
    \end{tabular}
    \label{tab:norm}
\end{table}

We set up two configurations that differ only in model size, and run SUAN and all variants A through G three times under each configuration, observing their AUC discrepancies, where variant F replaces RMSNorm with standard Layer-Norm, and variant G replaces the pre-norm strategy with the post-norm strategy.

We find that only variants F and G with a model size of size-4 exhibit performance degradation, as indicated by their mean AUCs and standard deviations shown in Table~\ref{tab:norm}. Their performance undermines the scaling law. We attribute these results to RMSNorm using the root mean square for layer normalization, which better reduces the impact of outliers and mitigates abnormal gradients, maintaining training stability. Further, pre-norm applies normalization at the start of the UAB, reducing numerical instability from subsequent complex operations. Conversely, post-norm normalizes after all UAB operations, which can easily cause gradient vanishing or explosion in deep models. 

Based on these findings, we reasonably speculate that the RMSNorm and pre-norm strategies are beneficial for maintaining the scaling law, preventing outliers and fluctuations from destroying the scaling laws.

\subsection{Deployment Optimization Analysis}

\begin{figure}[t]
     \centering
     \begin{subfigure}{.23\textwidth}
         \centering
         \includegraphics[width=\linewidth]{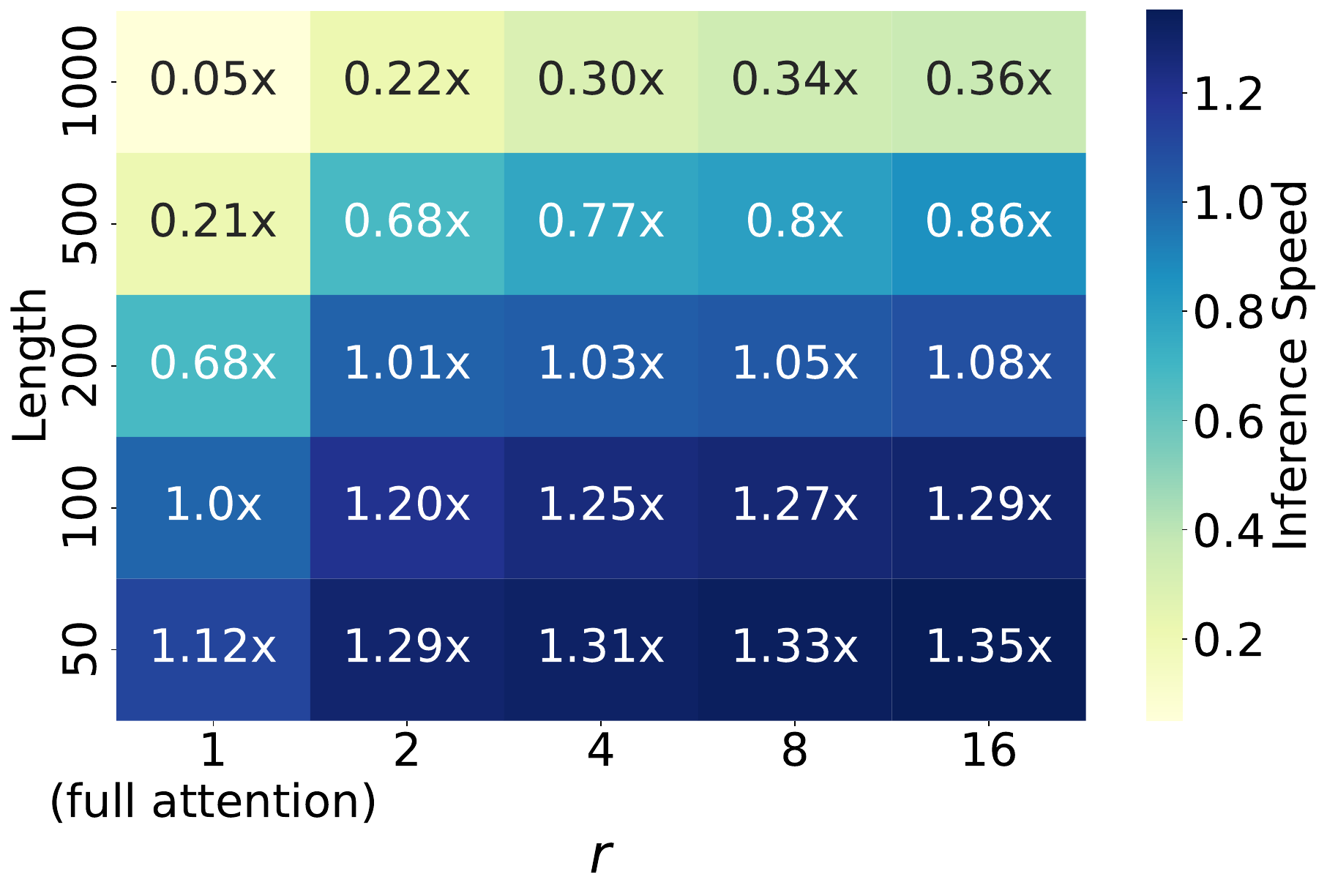}
         \caption{}
         \label{fig:sparse_efficiency}
     \end{subfigure} 
     \begin{subfigure}{.23\textwidth}
         \centering
         \includegraphics[width=\linewidth]{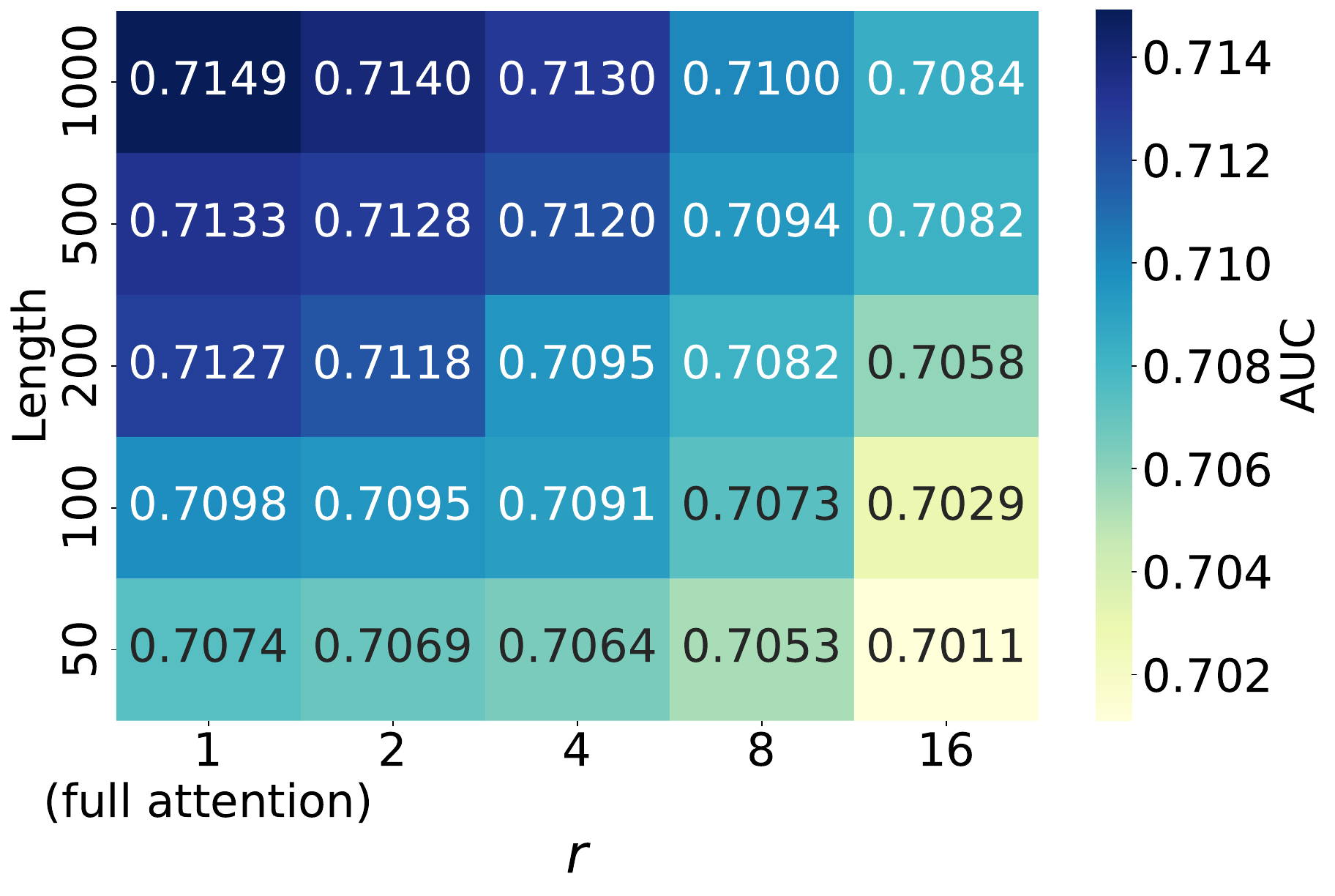}
         \caption{}
         \label{fig:sparse_performance}
     \end{subfigure}
     \caption{The impact of sparse self-attention on the model's (a) Efficiency, and (b) Performance.}
     \label{fig:sparse_impact}
\vspace{-10pt}
\end{figure}

\subsubsection{The impact of sparse self-attention.}
We choose the LightSUAN whose model size is set to size-2 and window size $k$ to 2, and observe inference speeds and AUCs on the industrial dataset across different dilated rates $r$ and sequence lengths, where the LightSUAN with $r=1$ is degenerated into the original SUAN with full attention. 

Figure~\ref{fig:sparse_efficiency} shows the relative inference speed with respect to the unit inference speed across varying dilated rates and sequence lengths, where the unit inference speed is calculated based on the time spent in inference for the model with full attention and a sequence length of 100. The results in Figure~\ref{fig:sparse_efficiency} illustrate that inference speed notably increases as the dilated rate increases. 
Figure~\ref{fig:sparse_performance} gives the AUC values under different dilated rates and sequence lengths. From the results in Figure~\ref{fig:sparse_performance}, it can be seen that there is no significant performance drop when $r<=4$ and AUC boosts rapidly as the sequence length grows longer.

\subsubsection{The impact of online distillation}
We first train one deployable low-grade LightSUAN and three high-performance high-grade SUAN models that are unable to be deployed online due to their long inference times. Table~\ref{table:online distillation1} details their configurations and AUCs on the industrial dataset. We then adopt online distillation to obtain three distilled LightSUAN models. Table~\ref{table:online distillation2} shows the distilled models and their AUCs on the industrial dataset.

The results reveal that distilled models can obtain enhanced performance. Notably, DistilSUAN-3 outperforms LightSUAN-1 by 6‰ and also exceeds the performance of both SUAN-1, which is equipped with a larger model size, and SUAN(L), which takes a longer sequence as input, as shown in Table~\ref{tab:performance_comparison}.

\begin{table}
\centering
\caption{Details of models. More $\star$s indicate a higher grade.}
\begin{tabular}{c|cccl|c}
\hline
Model       & Model Size   & Length       & Grade & Sparse   & AUC    \\
\hline
LightSUAN-1 & size-2        & 100       & $\star$ & $r$=2, $k$=2    & 0.7095 \\ 
SUAN-1      & size-4        & 100       & $\star\star$ & -    & 0.7133 \\ 
SUAN-2      & size-4        & 200       & $\star\!\star\!\star$  & -   & 0.7146 \\ 
SUAN-3      & size-5        & 1000      & $\star\!\star\!\star\star$ & -    & 0.7168 \\ 
\hline
\end{tabular}
\label{table:online distillation1}
\end{table}

\begin{table}
\centering
\caption{Performance comparison of distilled models.}
\begin{tabular}{c|cc|c}
\hline
\multirow{2}{*}{\shortstack{Distilled \\ Model}} & \multirow{2}{*}{\shortstack{Low-grade Model \\(as student)}} & \multirow{2}{*}{\shortstack{High-grade  Model\\(as teacher)}} & \multirow{2}{*}{AUC}  \\
 & & & \\  
\hline
DistilSUAN-1 & \multirow{3}{*}{LightSUAN-1} & SUAN-1 & 0.7114 \\ 
DistilSUAN-2 &  & SUAN-2 & 0.7122 \\ 
DistilSUAN-3 &  & SUAN-3 & 0.7139 \\ 
\hline
\end{tabular}
\label{table:online distillation2}
\end{table}

\subsection{Online A/B Test}
We conducted an online A/B test that lasted for three weeks. Considering the limited computational budget and the constraint on  online inference time, we deploy the original SUAN with a model size of size-2 and sequence length of 100, and DistilSUAN-3 in the live production environment of an online service, providing list advertisement recommendations. The baseline is the original online-serving CTR model, which has undergone several rounds of model upgrades, getting the essence from multiple basic models, including DIN, CAN, SIM, and other multi-behavior models.

Compared to the baseline, the original SUAN improves CTR by 2.67\% and CPM by 1.63\%, with an increase in average inference time from 33ms to 48ms, and the distilled LightSUAN enhances CTR by 2.81\% and CPM by 1.69\%, achieving this with an average inference time of 43ms.


\section{Conclusion}
\label{Conclusion}
This paper explores the potential of modeling user behaviors to make CTR prediction performance conform to the scaling laws and presents a knowledge distillation approach to allowing an online service to benefit from a CTR model with scaling laws. This study not only offers new insights into CTR predictions to support online performance improvement in the industry but also shares hands-on experience in identifying factors (e.g., model grade) and components (e.g., regularizers) affecting scaling laws of the CTR model. 

\begin{acks} 
This work was supported by the National Natural Science Foundation of China under Grant No. 62072450 and Meituan.
\end{acks}

\bibliographystyle{ACM-Reference-Format}
\bibliography{sample-base}

\end{document}